\documentclass[aps,amsmath,prd,nofootinbib,reprint,longbibliography]{revtex4-2}
\usepackage{graphicx}% Include figure files
\usepackage{dcolumn}% Align table columns on decimal point
\usepackage{bm,amsmath,amssymb,bbold}% bold math
\usepackage{verbatim}
\usepackage{dsfont,BOONDOX-cal}
\usepackage{wrapfig}
\usepackage{lipsum}
\usepackage{natbib}
\usepackage{BOONDOX-frak}
\usepackage[mathlines]{lineno}% Enable numbering of text and display math
\usepackage{lipsum}
\usepackage[colorlinks=true,urlcolor=blue,linkcolor=red,citecolor=blue]{hyperref}
\begin{document}
\title{Entanglement degradation as a tool to detect signatures of quantum gravity}% Force line breaks with \\
%\thanks{A footnote to the article title}%

\author{Soham Sen}
\email{sensohomhary@gmail.com, soham.sen@bose.res.in}

\author{Arnab Mukherjee}
\email{mukherji.arn@gmail.com}

\author{Sunandan Gangopadhyay}
\email{sunandan.gangopadhyay@gmail.com}

\affiliation{Department of Astrophysics and High Energy Physics\\
S. N. Bose National Centre for Basic Sciences, JD Block, Sector-III, Salt Lake, Kolkata 700106, India}

\begin{abstract}
\noindent We investigate entanglement degradation in the vicinity of a quantum corrected black hole. We consider a biprtite system (Alice-Rob) with Alice freely falling (radially) into the event horizon of a quantum corrected black hole and Rob being in the vicinity of the event horizon of the black hole. We consider a maximally entangled state (in the Fock basis) and start with the basic assumption that Rob is an uniformly accelerated observer. We then give a pedagogical analysis of the relation involving the Minkowski vaccum state and Rindler number states. Following the analogy given in \href{https://link.aps.org/doi/10.1103/PhysRevD.82.064006}{Phys. Rev. D 82 (2010) 064006}, we establish the relation between the Hartle-Hawking vacuum state and Boulware and Anti-Bouware number states from the Minkowski-Rindler relation. We then write down the quantum corrected black hole metric by making use of the near horizon approximation in an appropriate form. Next, we obtain the analytical forms of logarithmic negativity and mutual information and plot as a function of Rob's distance from the $r=0$ point. We observe that the entanglement degradation slows down due to the incorporation of quantum gravity corrections in the Schwarzschild black hole. This observation may lead to identification of quantum gravity signatures in future generation of advanced observational scenarios. We can also interpret this effect as a noisy quantum channel with an operator sum representation of a completely positive and trace preserving (CPTP) map. We then finally obtain the entanglement fidelity using this operator sum representation.
\end{abstract}
\maketitle

\section{Introduction}
\noindent Our day to day classical information theory is restricted to a binary system where we can make use of $0$ and $1$ as measures of information stored or communicated. With the advent of quantum mechanics in the first quarter of twentieth century, the idea of a quantum version of the classical information theory came as a byproduct of the quantum superposition principle. This new branch of physics was later named as quantum information theory. The relativistic generalization of quantum information theory, which involves general relativity, quantum field theory and quantum information theory, is also known as relativistic quantum information theory. The study of quantum correlations in case of a noninertial perspective is a very intersting sector in the genre of relativistic quantum information technology \cite{AlsingMilburn,TerashimaUeda,Shi,Ivette_Fuentes,
Alsing2,Alsing3,Ball,Adesso,Bradler,Ling,Ahn,Pan,
Doukas,VerSteeg,Leon,Adesso2,Datta,Lin,Wang,Martinez,
Alsing4,Ahn2}. In several of these works, the case of an entangled bipartite system was investigated when one of the observer was uniformly accelerated. The idea was to transport the stationary state to the Rindler space in order to truly investigate the effect of acceleration. In all of these cases the entangled states were taken as Fock states and instead of entanglement between spins, entanglement between number states were considered. In \cite{Ivette_Fuentes}, the generic Alice-Rob picture in the Minkowski-Rindler background was transferred to the black hole picture for bosonic fields. This study was inadequate in a sense that the Rindler horizon and the event horizon of a Schwarzschild black hole are very different in nature. The Rindler horizon can only be perceived by an accelerated observer whereas the event horizon exists for all observers. To deal with this problem, in \cite{Martinez} a one to one correspondance was observed among different vacuums from both the Minkowski and curved spacetimes. The system consists of two observers, Alice and Rob. Alice is freely falling into the event horizon of a Schwarzschild black hole and Rob is at a fixed radial distance just outside the event horizon of the black hole. Both Alice and Rob are observing a bipartite quantum state and the state is maximally entangled for the freely falling observer. Rob sees a degradation in the state due to the Hawking effect. In their analysis it was shown that the major interesting entanglement behaviours are observed in the vicinity of the event horizon. In case of the black hole picture when the observer is on the event horizon of the black hole, it imitates the infinite acceleration case in the Rindler spacetime.

\noindent In our analysis, we shall consider Alice to be freely falling into the event horizon of a quantum corrected black hole and Rob to be at a fixed distance just outside the event horizon of the same. Our main motivation behind this analysis is to investigate the effects of quantum gravitational corrections on entanglement degradation. The line element of the quantum corrected black hole spacetime following from a renormalization group approach of gravity is given by \cite{BonanoReuter}
\begin{equation}\label{0.1}
ds^2=-f(r)dt^2+\frac{1}{f(r)}dr^2+r^2d\Omega^2
\end{equation}
where 
\begin{equation}\label{0.2}
f(r)=1-\frac{2G(r)M}{r}
\end{equation}
with 
\begin{equation}\label{0.3}
G(r)=\frac{G }{1+\frac{\tilde{\omega} G}{r^2}}~.
\end{equation}
Throughout our analysis, we have used $\hbar=c=1$. The metric structure, given above, originates from the well known ``\textit{asymptotic safety approach}" to quantum gravity. This formalism revolves around an effective average action and by taking into consideration all loop effects, this effective action describes all gravitational phenomenon \cite{Reuter2, Wetterich1, Reuter3Wetterich2, Saueressig}. This action satisfies a renormalization group equation which results in the flow of the Newton's gravitational constant as a function of this scale. Using the flow of the Newton's gravitational constant, the metric in eq.(\ref{0.1}) was obtained where the constant $\tilde{\omega}$ carries quantum gravity corrections to the black hole geometry as a result of this renormalization group approach. 

\noindent At first, we have used the near horizon approximation to probe any static spherically symmetric black hole metric in the well known Rindler form and following the analysis in \cite{Martinez}, we have then obtained three unique time-like Killing vectors. The positive frequency modes associated with these Killing vectors let one define three unique vacuum states (Hartle-Hawking, Boulware and anti-Boulware).  Finally, one can obtain the relation between the Hartle-Hawking vacuum state and  Boulware - anti-Boulware Fock space basis. Using this relation, we then calculate the logarithmic negativity and mutual information for the reduced density matrix (where all the anti-Boulware states have been traced out). For the next part of our analysis, we have used the formalism in \cite{Ahn2} and shown that the entanglement degradation due to the Hawking effect can be described via a quantum channel with a completely positive and trace preserving  (CPTP) map. We then finally compute the entanglement fidelity to investigate how the quantum channel preserves the initial entanglement between the two parties of the bipartite state. It is important to note that we have considered only bosonic field modes in our analysis.

\noindent The construction of the paper goes as follows. In section section (\ref{II}), we give a brief preview of the Alice-Rob system and obtain the relation between the Minkowski vacuum and Rindler Fock state basis. In section (\ref{III}), we express a static spherically symmetric black hole in the Rindler form and obtain the analogy between several vacuum states. In section (\ref{IV}), we obtain the analytical forms of the logarithmic negativity and mutual information for a quantum corrected black hole and plotted against the distance of the observer from the $\mathcal{r}=0$ point. In section (\ref{V}), we investigate the entire process as a quantum channel with a CPTP map and obtain the analytical form of the entanglement fidelity for a quantum corrected black hole. Finally, we conclude our analysis in section (\ref{VI}).
\section{Minkowski-Rindler identification: A brief review}\label{II}
\noindent In this section, we start by providing a detailed and pedagogical derivation of the expression connecting Minkowski vacuum state and the product of two mode squeezed states of the Rindler vacuum \cite{Ivette_Fuentes}. The Rindler coordinate system is the one describing an uniformly accelerated observer.  The Minkowski coordinates in 3+1- spacetime dimensions is given by $\{t,x,y,z\}$ and the Rindler coordinates are denoted by $\{\bar{t},\bar{x},\bar{y},\bar{z}\}$. In region I (right Rindler wedge), we can express the Minkowski coordinates, in terms of the Rindler coordinates as
\begin{equation}\label{1.1}
\begin{split}
t&=\bar{z}\sinh a\bar{t},~x=\bar{x},~y=\bar{y},~z=\bar{z}\cosh a\bar{t}
\end{split}
\end{equation}
and in region IV (left Rindler wedge)
\begin{equation}\label{1.2}
t=-\bar{z}\sinh a\bar{t},~x=\bar{x},~y=\bar{y},~z=-\bar{z}\cosh a\bar{t}~.
\end{equation}
In order to interconnect the coordinate transformation between the two coordinate systems, we have considered that the observer is uniformly accelerating in the $z$ direction only with an uniform acceleration $a$. In order to proceed further, we now come down to a 1+1-dimensional analysis involving $(t,z)$ coordinates only. The massless Klein-Gordon equation for a scalar field in the Minkowski background reads 
\begin{equation}\label{1.3}
\partial_\mu\partial^\mu\phi(t,z)=0\implies(\partial_t^2-\partial_z^2)\phi(t,z)=0~.
\end{equation} 
In order to define the normalization constant, we need to write down the Lorentz invariant inner product, which is given by
\begin{equation}\label{1.4}
(\phi_1,\phi_2)=-i\int_{\Sigma}d\Sigma^\mu\left(-\phi_1^*\partial_\mu\phi_2+\phi_2\partial_\mu\phi_1^*\right)
\end{equation}
where $\Sigma$ is a spacelike hypersurface. Now for a constant time hypersurface, we can simplify the above inner product in the following form (in 1+1-dimensions)
\begin{equation}\label{1.5}
(\phi_1,\phi_2)=-i\int_{z}dz\left(-\phi_1^*\partial_t\phi_2+\phi_2\partial_t\phi_1^*\right)~.
\end{equation}
Using a separation of variables method, we can obtain a solution of the Klein-Gordon equation (eq.(\ref{1.3})) and write down the analytical forms of the Minkowski field modes as 
\begin{equation}\label{1.6}
u_k^{\mathcal{M}}(t,z)=\mathcal{N}(k)e^{-i\omega t+i \omega z}
\end{equation}
where $\omega=k$ when the speed of light is set equal to unity and $\mathcal{N}(k)(=\mathcal{N}_\omega)$ is a real and undetermined normalization constant. We shall now make use of eq.(\ref{1.5}) to determine the undetermined normalization constant given by
\begin{equation}\label{1.7}
\begin{split}
(u_k^{\mathcal{M}},u_{k'}^{\mathcal{M}})&=-i\int_{-\infty}^\infty dz\left[-{u_k^{\mathcal{M}}}^*\partial_t u_{k'}^{\mathcal{M}}+u_{k'}^{\mathcal{M}}\partial_t{u_k^{\mathcal{M}}}^*\right]\\
\implies\delta(\omega-\omega')&=2\pi(\omega+\omega')\mathcal{N}_\omega\mathcal{N}_{\omega'} \delta(\omega-\omega')\\
&=4\pi\omega\mathcal{N}_\omega^2\delta(\omega-\omega')\\
\implies \mathcal{N}_\omega&=\frac{1}{\sqrt{4\pi\omega}}~.
\end{split}
\end{equation}
Using the above form of the normalization constant,  we can finally write down the Minkowski mode solution as
\begin{equation}\label{1.8}
u_k^{\mathcal{M}}(t,z)=\frac{1}{\sqrt{4\pi\omega}}e^{-i\omega t+i \omega z}.
\end{equation}
Next, we shall be calculating the Rindler modes in region I. We start by obtaining the Klein-Gordon equation in Rindler coordinates. To do this we write down the relations among the partial derivatives corresponding to Minkowski and Rindler coordinates:
\begin{align}
\partial_t&=\frac{\partial \bar{t}}{\partial t}\partial_{\bar{t}}+\frac{\partial \bar{z}}{\partial t}\partial_{\bar{z}}\nonumber\\&=\frac{1}{a\bar{z}}\cosh a\bar{t}\partial_{\bar{t}}-\sinh a\bar{t}\partial_{\bar{z}}~,\label{1.9}\\
\partial_z&=\frac{\partial \bar{t}}{\partial z}\partial_{\bar{t}}+\frac{\partial \bar{z}}{\partial z}\partial_{\bar{z}}\nonumber\\&=-\frac{1}{a\bar{z}}\sinh a\bar{t}\partial_{\bar{t}}+\cosh a\bar{t}\partial_{\bar{z}}~.\label{1.10}
\end{align}
Using eq.(s)(\ref{1.9},\ref{1.10}) back in eq.(\ref{1.3}), we obtain the Klein-Gordon equation in the Rindler spacetime to be
\begin{equation}\label{1.11}
\begin{split}
(\partial_t^2-\partial_z^2)\phi(t,z)=\frac{1}{a^2\bar{z}^2}\left(\partial_{\bar{t}}^2-a^2\partial_{\ln \bar{z}}^2\right)\phi(\bar{t},\bar{z})=0~.
\end{split}
\end{equation}
Solving eq.(\ref{1.11}) and making use of the inner product definition (eq.(\ref{1.5})), we obtain the Rindler mode solution in region I  to be
\begin{equation}\label{1.12}
u_{k,\pm}^{\mathcal{R}_I}=\frac{1}{\sqrt{4\pi\omega}}e^{-i\omega\bar{t}\pm\frac{i\omega}{a}\ln\bar{z}}~.
\end{equation} 
In this analysis, we shall be mainly considering the $u_{k,+}^{\mathcal{R}_I}$ mode solutions. In terms of the Minkowski coordinates, the Rindler mode solution in eq.(\ref{1.12}) reads
\begin{equation}\label{1.13}
u_{k,\pm}^{\mathcal{R}_I}=\sqrt{\frac{a}{4\pi\omega}}\left(\frac{z\mp t}{l_\omega}\right)^{\pm\frac{i\omega}{a}}=\frac{1}{\sqrt{4\pi\Omega}}\left(\frac{z\mp t}{l_\Omega}\right)^{\pm i\Omega}\equiv u_{\Omega,\pm}^{I}
\end{equation}
where $\Omega$ $(=\frac{\omega}{a})$ is a dimensionless constant, $l_\omega=l_\Omega$ has dimension of length in natural units and $u_{\Omega,+}^I$ denotes field modes which are propagating to the right direction along lines of constant $z-t$. Similarly, the Rindler mode solutions in region IV reads
\begin{equation}\label{1.14}
u_{k,\pm}^{\mathcal{R}_{IV}}=\frac{1}{\sqrt{4\pi\Omega}}\left(\frac{\pm t- z}{l_\Omega}\right)^{\mp i\Omega}\equiv u_{\Omega,\pm}^{IV}~.
\end{equation}
As we shall mainly be considering the right moving modes, we shall be omitting the plus sign while writing down the mode solutions. 

\noindent One can now do a second quantization of the classical field $\phi$ which satisfies the Klein-Gordon equation given by $\square \hat{\phi}=0$. In terms of the Minkowski mode solutions and the corresponding creation and annihilation operators, we can write down the quantized scalar field as 
\begin{equation}\label{1.15}
\hat{\phi}=\int dk \left(u_{k}^\mathcal{M}(t,z)\hat{a}_{k,\mathcal{M}}+{u_{k}^\mathcal{M}}^*(t,z)\hat{a}^\dagger_{k,\mathcal{M}}\right)
\end{equation} 
where the creation and annihilation operators satisfy the following commutation relation
\begin{equation}\label{1.16}
[\hat{a}_{k,\mathcal{M}},\hat{a}_{k',\mathcal{M}}^\dagger]=\delta(k-k')~.
\end{equation}
The action of the annihilation operator on the vacuum state corresponding to a fixed field mode is defined as
\begin{equation}\label{1.17}
\hat{a}_{k,\mathcal{M}}|0\rangle^k_{\mathcal{M}}=0
\end{equation}
and the total Minkowski vacuum state is defined as a product of all the individual vacuum states corrponding to each field modes as
\begin{equation}\label{1.18}
|0\rangle_\mathcal{M}=\prod_{k}|0\rangle^k_{\mathcal{M}}~.
\end{equation}
It is important to note that the mode solutions in regions I and IV provide a complete set of orthonomal solutions. As a result, one can express the field $\hat{\phi}$ in terms of the Rindler mode solutions as 
\begin{equation}\label{1.19}
\hat{\phi}=\int d\Omega \left(u_{\Omega}^{I}\hat{a}_{\Omega,I}+{u_{\Omega}^{I}}^*\hat{a}^\dagger_{\Omega,I}+u_{\Omega}^{IV}\hat{a}_{\Omega,IV}+{u_{\Omega}^{IV}}^*\hat{a}^\dagger_{\Omega,IV}\right)
\end{equation}
where the creation and the annihilation operators act on the vacuum states of the two Rindler wedges respectively as
\begin{align}
\hat{a}_{\Omega,I}\otimes\mathbb{1}_{IV}|0_I,0_{IV}\rangle&=\left(\hat{a}_{\Omega,I}|0_I\rangle\right)\otimes\left(\mathbb{1}_{IV}|0_{IV}\rangle\right)=0~,\label{1.20}\\
\mathbb{1}_{I}\otimes\hat{a}_{\Omega,IV}|0_I,0_{IV}\rangle&=\left(\mathbb{1}_{I}|0_{I}\rangle\right)\otimes\left(\hat{a}_{\Omega,IV}|0_IV\rangle\right)=0~.\label{1.21}
\end{align}
It is to be noted that region I and region IV are causally disconnected and as a result it is possible to write down the following commutation relations
\begin{align}
[\hat{a}_{\Omega,I},\hat{a}^{\dagger}_{\Omega',I}]&=[\hat{a}_{\Omega,IV},\hat{a}^{\dagger}_{\Omega',IV}]=\delta(\Omega-\Omega')~,\label{1.22}\\
[\hat{a}_{\Omega,I},\hat{a}_{\Omega',I}]&=[\hat{a}^\dagger_{\Omega,I},\hat{a}^\dagger_{\Omega',I}]=[\hat{a}_{\Omega,I},\hat{a}^\dagger_{\Omega',IV}]=0~,\label{1.23}\\
[\hat{a}_{\Omega,IV},\hat{a}_{\Omega',IV}]&=[\hat{a}^\dagger_{\Omega,IV},\hat{a}^\dagger_{\Omega',IV}]=[\hat{a}^\dagger_{\Omega,I},\hat{a}_{\Omega',IV}]=0~.\label{1.24}
\end{align}
We now need to express the creation and annihilation operators of the Minkowski states in terms of the creation and annihilation operators of the Rindler states. Before proceeding with this analysis, we need to remember that the mode solutions $u_{k}^\mathcal{M}$ satisfies the following relation with respect to the inner product defined in eq.(\ref{1.5}) as
\begin{equation}\label{1.25}
(u_k^\mathcal{M},u_{k'}^\mathcal{M})=\delta(k-k')~,~~(u_k^\mathcal{M},{u_{k'}^\mathcal{M}}^*)=0~.
\end{equation}
Taking the inner product of $\hat{\phi}$ (for the decomposition of $\hat{\phi}$ in terms of Minkowski field modes) with $u_{k'}^{\mathcal{M}}$, we obtain the following relation
\begin{equation}\label{1.26}
\begin{split}
\left(u_{k'}^\mathcal{M},\hat{\phi}\right)&=\int dk\left((u_{k'}^\mathcal{M},u_{k}^\mathcal{M})\hat{a}_{k,\mathcal{M}}+(u_{k'}^\mathcal{M},{u_{k}^\mathcal{M}}^*)\hat{a}^\dagger_{k,\mathcal{M}}\right)\\
&=\int dk \delta(k-k')\hat{a}_{k,\mathcal{M}}\\
&=\hat{a}_{k',\mathcal{M}}~.
\end{split}
\end{equation}
It is now possible to substitute the mode expansion of $\hat{\phi}$ from eq.(\ref{1.19}) in the left hand side of the above equation and we can recast eq.(\ref{1.26}) in the following form
\begin{equation}\label{1.27}
\begin{split}
\hat{a}_{k,\mathcal{M}}=&\int d\Omega \Bigr((u_{k}^\mathcal{M},u_\Omega^I)\hat{a}_{\Omega,I}+(u_{k}^\mathcal{M},{u_\Omega^I}^*)\hat{a}^\dagger_{\Omega,I}\\
&+(u_{k}^\mathcal{M},u_\Omega^{IV})\hat{a}_{\Omega,IV}+(u_{k}^\mathcal{M},{u_\Omega^{IV}}^*)\hat{a}^\dagger_{\Omega,IV}\Bigr)~.
\end{split}
\end{equation}
We shall now evaluate all of the four innerproducts in the above equation. Applying the definition of the inner product, the first inner product turms out to be
\begin{equation}\label{1.28}
\begin{split}
(u_{k}^\mathcal{M},u_\Omega^I)&=-i\int dz\left({u_{k}^{\mathcal{M}}}^*\partial_tu_{\Omega}^I+u^I_{\Omega}\partial_t{u_k^\mathcal{M}}^*\right)\\
&=\frac{1}{4\pi l_\Omega^{i\Omega}\sqrt{\omega\Omega}}\int dz\biggr(\Omega (z-t)^{i\Omega-1}\\&+\omega (z-t)^{i\Omega}\biggr)e^{-i\omega(z-t)}~.
\end{split}
\end{equation} 
We shall now make a change of coordinates given by $z-t=\zeta$ and in the Rindler wedge I, $(z-t)>0$. Hence, $\zeta$ will range from 0 to $\infty$. We can recast eq.(\ref{1.28}) in the following form
\begin{equation}\label{1.29}
\begin{split}
(u_{k}^\mathcal{M},u_\Omega^I)&=\frac{1}{4\pi l_\Omega^{i\Omega}\sqrt{\omega\Omega}}\int_0^\infty d\zeta \left(\Omega\zeta^{i\Omega-1}+\omega\zeta^{i\Omega}\right)e^{-i\omega\zeta}\\
&=\frac{\Omega (i\omega)^{-i\Omega}}{2\pi l_\Omega^{i\Omega}\sqrt{\omega\Omega}}\Gamma[i\Omega]\\
&=\frac{\Omega(i l_\Omega\omega)^{-i\Omega}}{2\pi\sqrt{\omega \Omega}}\sqrt{\frac{\pi}{\Omega}}\sqrt{\frac{2}{e^{\pi\Omega}-e^{-\pi\Omega}}}e^{i\text{arg}[\Gamma[i\Omega]]}\\
&=\frac{1}{\sqrt{2\pi\omega}}(l_\Omega e^{-\frac{\phi}{\Omega}}\omega)^{-i\Omega}\frac{1}{\sqrt{1-e^{-2\pi\Omega}}}\\
&=\frac{1}{\sqrt{2\pi\omega}}(l\omega)^{-i\Omega}\frac{1}{\sqrt{1-e^{-2\pi\Omega}}}
\end{split}
\end{equation}
where $\phi\equiv\text{Arg}[\Gamma[i\Omega]]$, $l\equiv l_\Omega e^{-\frac{\phi}{\Omega}}$, and $(i)^{-i\Omega}=(e^{\frac{i\pi}{2}})^{-i\Omega}=e^{\frac{\pi\Omega}{2}}$. The next innerproduct of $u_k^\mathcal{M}$ with ${u_\Omega^I}^*$ is given as follows
\begin{equation}\label{1.30}
(u_k^\mathcal{M},{u_\Omega^I}^*)=-\frac{1}{\sqrt{2\pi\omega}}(l\omega)^{i\Omega}\frac{e^{-\pi\Omega}}{\sqrt{1-e^{-2\pi\Omega}}}~.
\end{equation}
The final two innerproducts have the forms given by
\begin{align}
(u_k^\mathcal{M},u_\Omega^{IV})&=\frac{1}{\sqrt{2\pi\omega}}(l\omega)^{i\Omega}\frac{1}{\sqrt{1-e^{-2\pi\Omega}}}~,\label{1.31}\\
(u_k^\mathcal{M},{u_\Omega^{IV}}^*)&=-\frac{1}{\sqrt{2\pi\omega}}(l\omega)^{-i\Omega}\frac{e^{-\pi\Omega}}{\sqrt{1-e^{-2\pi\Omega}}}\label{1.32}~.
\end{align}
With a new redefinition $e^{-\pi\Omega}\equiv\tanh r_\Omega$, we can recast eq.(\ref{1.27}) as 
\begin{align}\label{1.33}
\begin{split}
\hat{a}_{k,\mathcal{M}}&=\int_0^\infty d\Omega \biggr({\alpha_{\omega,\Omega}^R}^*\left(\cosh r_\Omega \hat{a}_{\Omega,I}-\sinh r_\Omega \hat{a}_{\Omega,IV}^\dagger\right)\\&+{\alpha_{\omega,\Omega}^L}^*\left(-\sinh r_\Omega\hat{a}^\dagger_{\Omega,I}+\cosh r_\Omega\hat{a}_{\Omega,IV}\right)\biggr)
\end{split}
\end{align}
where ${\alpha_{\omega,\Omega}^R}^*=\frac{1}{\sqrt{2\pi\omega}}(l\omega)^{-i\Omega}$ and ${\alpha_{\omega,\Omega}^L}^*=\frac{1}{\sqrt{2\pi\omega}}(l\omega)^{i\Omega}$ are the Bogoliubov coefficients. 

\noindent We can now express the right and left moving Unruh annihilation operators as
\begin{align}
\hat{a}^R_\Omega&=\cosh r_\Omega \hat{a}_{\Omega,I}-\sinh r_\Omega \hat{a}_{\Omega,IV}^\dagger~,\label{1.34}\\
\hat{a}^L_\Omega&=-\sinh r_\Omega\hat{a}^\dagger_{\Omega,I}+\cosh r_\Omega\hat{a}_{\Omega,IV}~.\label{1.35}
\end{align}
By means of eq.(s)(\ref{1.34},\ref{1.35}), we can indeed reexpress the Minkowski annihilation operator in eq.(\ref{1.33}) as 
\begin{equation}\label{1.36}
\hat{a}_{k,\mathcal{M}}=\int_0^\infty d\Omega \left({\alpha_{\omega,\Omega}^R}^*\hat{a}^R_\Omega+{\alpha_{\omega,\Omega}^L}^*\hat{a}^L_\Omega
\right)~.
\end{equation}
It is important to note from eq.(\ref{1.36}) that the Minkowski annihilation operator can be expressed as a combination of the Unruh annihilation operators only. As a result the Unruh annihilation operator will annihilate the Minkowski vacuum as well. Hence, we can write down the following relation
\begin{equation}\label{1.37}
\hat{a}_{\omega,\mathcal{M}}|0\rangle_\mathcal{M}=\hat{a}_{\Omega}^R|0\rangle_\mathcal{M}=\hat{a}_{\Omega}^L|0\rangle_\mathcal{M}=0~.
\end{equation}
From eq.(\ref{1.37}), it is straightforward to conclude that the Minkowski vacuum and Unruh vacuum are identical which can be represented in the following form
\begin{equation}\label{1.38}
|0\rangle_\mathcal{M}=|0\rangle_U=\prod_\Omega|0\rangle^\Omega_U
\end{equation}
where $|0\rangle^\Omega_U$ is the Unruh vacuum corresponding to an individual field mode with frequency $\Omega$. 

\noindent Now we take an ansatz given by
\begin{equation}\label{1.39}
\begin{split}
|0\rangle_U^\Omega=\sum_n f_\Omega(n)|n\rangle^\Omega_{I}\otimes|n\rangle^\Omega_{IV}
\end{split}
\end{equation}
where $f_\Omega(n)$ is an unknown normalization factor, dependent on the dimensionless number $\Omega$. Before acting with $\hat{a}_{\Omega}^R$ on the both sides of eq.(\ref{1.39}), we need to express $\hat{a}_\Omega^R$ rigorously as
\begin{equation}\label{1.40}
\begin{split}
\hat{a}^R_\Omega&=\cosh r_\Omega ~\hat{a}_{\Omega,I}\otimes \mathbb{1}_{IV}-\sinh r_\Omega ~\mathbb{1}_I\otimes\hat{a}_{\Omega,IV}^\dagger~.
\end{split}
\end{equation}
Action of $\hat{a}_\Omega^R$ from eq.(\ref{1.40}) on the both sides of eq.(\ref{1.39}) is given by
\begin{equation}\label{1.41}
\begin {split}
0=\hat{a}_\Omega^R|0\rangle_U^\Omega=&\sum_nf_\Omega(n)\Bigr(\cosh r_\Omega \hat{a}_{\Omega,I}|n\rangle_I^\Omega\otimes|n\rangle_{IV}^\Omega\\&-\sinh r_\Omega|n\rangle_I^\Omega\hat{a}^\dagger_{\Omega,IV}|n\rangle_{IV}^\Omega\Bigr)\\
=&\sum_nf_\Omega(n)\Bigr(\cosh r_\Omega\sqrt{n}|n-1\rangle^\Omega_I\otimes|n\rangle^\Omega_{IV}\\
&-\sinh r_\Omega\sqrt{n+1}|n\rangle_I^\Omega\otimes|n+1\rangle^\Omega_{IV}\Bigr)~.
\end{split}
\end{equation}
We now act with $~^\Omega_I\!\langle m|\otimes~^\Omega_{IV}\!\langle m'|$ from the left in the above equation and we can then recast eq.(\ref{1.41}) as
\begin{equation}\label{1.42}
\begin{split}
0=&\sum_nf_\Omega(n)\Bigr(\cosh r_\Omega \sqrt{n}~^\Omega_I\!\langle m|n-1\rangle_I^\Omega~^\Omega_{IV}\!\langle m'|n\rangle_{IV}^\Omega\\
&-\sinh r_\Omega\sqrt{n+1}~^\Omega_I\!\langle m|n\rangle_I^\Omega~^\Omega_{IV}\!\langle m'|n+1\rangle_{IV}^\Omega\Bigr)\\
=&\sum_nf_\Omega(n)\Bigr(\cosh r_\Omega\sqrt{n}\delta_{m,n-1}\delta_{m',n}\\&-\sinh r_\Omega\sqrt{n+1}\delta_{m,n}\delta_{m',n+1}\Bigr)\\
=&\left(f_\Omega(m')\cosh r_\Omega-f_\Omega(m'-1)\sinh r_\Omega\right)\sqrt{m'}\delta_{m',m+1}~.
\end{split}
\end{equation}
From eq.(\ref{1.42}), we can write down a recursion relation in $f_\Omega(n)$ as follows
\begin{equation}\label{1.43}
f_\Omega(n)=\tanh^n r_\Omega f_\Omega(0)~.
\end{equation}
We need to determine the constant $f_\Omega(0)$ by imposing the normalization condition of $|0\rangle_U^\Omega$  as follows
\begin{equation}\label{1.44}
\begin{split}
1&=~_U^\Omega\!\langle0|0\rangle_U^\Omega\\&=f_\Omega^2(0)\sum_{n,m}(\tanh r_\Omega)^{n+m}~^\Omega_I\!\langle m|n\rangle_I^\Omega~^\Omega_{IV}\!\langle m|n\rangle_{IV}^\Omega\\
&=f_\Omega^2(0)\sum_n\tanh^{2n}r_\Omega\\
&=f_\Omega^2(0)\cosh^2 r_\Omega\\
\implies f_\Omega(0)&=\frac{1}{\cosh r_\Omega}~.
\end{split}
\end{equation}
Using the form of $f_\Omega(n)$ and $f_\Omega(0)$ from eq.(s)(\ref{1.43},\ref{1.44}), we can recast eq.(\ref{1.39}) as
\begin{equation}\label{1.46}
|0\rangle_U^\Omega=\frac{1}{\cosh r_\Omega}\sum_n \tanh ^nr_\Omega|n\rangle^\Omega_{I}|n\rangle^\Omega_{IV}
\end{equation}
where for simplicity we have omitted the tensor product sign. The Unruh vacuum state and the Minkowski vacuum state coincides, hence from eq.(\ref{1.46}) we can write down the following relation
\begin{equation}\label{1.47}
|0\rangle_\mathcal{M}^k=\frac{1}{\cosh r_\Omega}\sum_n \tanh ^nr_\Omega|n\rangle^\Omega_{I}|n\rangle^\Omega_{IV}~.
\end{equation}
We shall also need to calculate the first excited state corresponding to the single mode Minkowski vacuum state. As the Minkowski and Unruh states can be mapped with one another, we start by calculating the first excited state in the Unruh vacuum corresponding to a single mode only. The complete Unruh raising operator is given by a linear combination of the raising operators corresponding to the left and right moving Unruh modes as follows
\begin{equation}\label{1.48}
\hat{a}_U^{\Omega\dagger}=\mathcal{A}_L\hat{a}_\Omega^{L\dagger}+\mathcal{A}_R\hat{a}_\Omega^{R\dagger}
\end{equation}
where $|\mathcal{A}_L|^2+|\mathcal{A}_R|^2=1$. A very convenient choice is to take $\mathcal{A}_R=1$ and $\mathcal{A}_L=0$.  We can hence write down the first excited state by using the operator action of $\hat{a}_U^{\Omega\dagger}$ on $|0\rangle_U^\Omega$ along with the determination of an appropriate normalization constant (following the earlier procedure) as
\begin{equation}\label{1.49}
|1\rangle_U^\Omega=|1\rangle_\mathcal{M}^k=\sum_n\frac{\sqrt{n+1}}{\cosh^2r_\Omega}\tanh^nr_\Omega|n+1\rangle_I^\Omega|n\rangle_{IV}^\Omega~.
\end{equation}
With eq.(s)(\ref{1.47},\ref{1.49}) in hand, we can now move towards the analysis of entangled Fock states in a curved background. 
\section{Near horizon analysis and the Rindler-Kruskal identification}\label{III}
\noindent In this section, we shall consider the quantum corrected black hole geometry and apply the near horizon approximation to recast the metric in a form which will help us to use the usual quantum information theoretic wisdom to analyze the entanglement degradation for a maximally entangled state on this black hole geometry. In this section, we shall follow the analysis used in \cite{Martinez}. 

\noindent The  line element for a static spherically symmetric black hole geometry with a lapse function $f(r)$ in 3+1-spacetime dimensions is given by
\begin{equation}\label{1.50}
ds^2=-f(r)dt^2+\frac{1}{f(r)}dr^2+r^2d\Omega^2
\end{equation}
where we have used the $\{-,+,+,+\}$ signature for the metric. We shall now make use of the near horizon approximation and write down the lapse function in the following form
\begin{equation}\label{1.51}
f(r)\simeq (r-r_+)f'(r_+)
\end{equation}
where $r_+$ is the event horizon radius of the black hole. We now make a change of coordinates given by
\begin{equation}\label{1.52}
\zeta=2\sqrt{\frac{r-r+}{f'(r_+)}}\implies r-r_+=\frac{\zeta^2}{4}f'(r_+)~.
\end{equation}
Using eq.(s)(\ref{1.51},\ref{1.52}), we can recast the line element in eq.(\ref{1.50}) in the following form (in 1+1-spacetime dimensions)
\begin{equation}\label{1.53}
ds^2=-\frac{\zeta^2{f'}^{2}(r_+)}{4}dt^2+d\zeta^2~.
\end{equation}
In terms of the surface gravity $\kappa=\frac{f'(r_+)}{2}$, we can recast the above equation in the following form
\begin{equation}\label{1.54}
ds^2=-\kappa^2\zeta^2dt^2+d\zeta^2~.
\end{equation}
We now consider an observer sitting at a distance $\mathcal{r}$ where the proper time of the observer is denoted by $\tau$. We can then write down the following relation
\begin{equation}\label{1.55}
\begin{split}
-d\tau^2&=-f(r)\rvert_{r=\mathcal{r}}dt^2+\frac{1}{f(r)}\biggr\rvert_{r=\mathcal{r}}d\mathcal{r}^2=-f(\mathcal{r})dt^2\\
\text{or, } \frac{dt}{d\tau}&=\frac{1}{\sqrt{f(\mathcal{r})}}\implies t=\frac{\tau}{\sqrt{f(\mathcal{r})}}~.
\end{split}
\end{equation}
In terms of the proper time $\tau$, we can recast eq.(\ref{1.54}) as follows
\begin{equation}\label{1.56}
ds^2=-\frac{\kappa^2\zeta^2}{f(\mathcal{r})}d\tau^2+d\zeta^2~.
\end{equation}
Now the value of the proper acceleration for an accelerated observer at some $r$ is defined as
\begin{equation}\label{1.57}
a=\sqrt{a_\mu a^\mu}
\end{equation}
where $a^\mu=\frac{\xi^\beta}{|\xi|}\nabla_\beta\left(\frac{\xi^\mu}{|\xi|}\right)$ gives the four acceleration with $v^\mu=\frac{\xi^\mu}{|\xi|}$ denoting the four velocity of the observer and $\xi^\mu=\{1,0,0,0\}$ being a timelike Killing vector. It is now straightforward to evaluate the four acceleration of the observer
\begin{equation}\label{1.58}
a^\mu=\left\{0,\frac{1}{2}\partial_rf,0,0\right\}~,~~a_\mu=g_{\kappa\mu}a^\kappa=\left\{0,\frac{1}{2f}\partial_rf,0,0\right\}~.
\end{equation}
Using eq.(\ref{1.58}), we can obtain the proper acceleration of the observer to be of  the form
\begin{equation}\label{1.59}
a(r)=\sqrt{a_\mu a^\mu}=\frac{\partial_r f}{2\sqrt{f(r)}}~.
\end{equation}
From eq.(\ref{1.59}), it is straightforward to infer that the acceleration becomes infinite when $r=r_+$.
In the near horizon approximation, we can evaluate the following relation
\begin{equation}\label{1.59b}
\partial_r f(r)\simeq\partial_r\left((r-r_+)f'(r_+)\right)=f'(r_+)=2\kappa~.
\end{equation} 
Hence, we can write down the proper acceleration for an observer sitting at a distance $\mathcal{r}$ from the $r=0$ point to be
\begin{equation}\label{1.60}
\mathcal{a}=a(\mathcal{r})=\frac{\kappa}{\sqrt{f(\mathcal{r})}}~.
\end{equation}
Using eq.(\ref{1.60}), we can recast eq.(\ref{1.56}) as 
\begin{equation}\label{1.61}
ds^2=-\mathcal{a}^2\zeta^2 d\tau^2+d\zeta^2~.
\end{equation}
Eq.(\ref{1.60}) depicts the fact that  any static spherically symmetric blackhole metric can be expressed in the Rindler form by means of near horizon approximation and the constant acceleration in the Rindler case is now replaced by the proper acceleration of an observer sitting at a fixed radial distance outside but in the vicinity of the event horizon of the black hole. 

\noindent Our next aim is to define timelike vectors. We start by writing down the null Kruskal-Szekeres coordinates as
\begin{align}
\mathcal{u}=-\frac{1}{\kappa}e^{-\kappa\left(t-\int\frac{dr}{f(r)}\right)}~,~~\mathcal{v}=\frac{1}{\kappa}e^{\kappa\left(t+\int\frac{dr}{f(r)}\right)}~.\label{1.62}
\end{align}
Using eq.(\ref{1.62}), one can write down the radial part of the black hole metric in the following form
\begin{equation}\label{1.63}
ds^2=-f(r)e^{-2\kappa\int\frac{dr}{f(r)}}d\mathcal{u}d\mathcal{v}~.
\end{equation}
Very near the horizon, eq.(\ref{1.63}) can be expressed as (keeping only leading constant terms and setting the integration constant to $\frac{1}{2\kappa}$) 
\begin{equation}\label{1.64}
ds^2\simeq-e^{-1}d\mathcal{u}d\mathcal{v}~.
\end{equation}
This analysis shows (following \cite{Martinez}) that there are three possible timelike Killing vectors. The first timelike Killing vector is $\partial_{\bar{t}}\propto\partial_\mathcal{u}+\partial_\mathcal{v}$, where this timelike vector is similar to the timelike Killing vector in the Minkowski spacetime. One can construct a vacuum state out of positive frequency modes associated with this timelike Killing vector and this vacuum state is also known as the Hartle-Hawking vacuum state. As a result of the analogy between the Killing vectors, we can also claim that the Hartle-Hawking vacuum state is analogous to the  Minkowski vacuum state. The Hawking-Hartle vacuum state is generally written as $|0\rangle_H$ and $|0\rangle_H\leftrightarrow|0\rangle_\mathcal{M}$. The second Killing vector is $\partial_t$ and it is straight forward to obtain a relation in terms of the $\{\mathcal{u},\mathcal{v}\}$ coordinate system as follows
\begin{equation}\label{1.65}
\begin{split}
\partial_t&=\frac{\partial \mathcal{u}}{\partial t}\frac{\partial}{\partial\mathcal{u}}+\frac{\partial \mathcal{v}}{\partial t}\frac{\partial}{\partial\mathcal{v}}\\
&=(-\kappa)\left[-\frac{1}{\kappa}e^{-\kappa\left(t-\int\frac{dr}{f(r)}\right)}\right]\frac{\partial}{\partial \mathcal{u}}+\kappa\left[\frac{1}{\kappa}e^{\kappa\left(t+\frac{dr}{f(r)}\right)}\right]\frac{\partial}{\partial\mathcal{v}}\\
&=-\kappa\left(\mathcal{u}\partial_\mathcal{u}-\mathcal{v}\partial_\mathcal{v}\right)~.
\end{split}
\end{equation}
From the above calculation, we deduce that $\partial_t\propto\mathcal{u}\partial_\mathcal{u}-\mathcal{v}\partial_\mathcal{v}$. $\partial_t$ is a timelike Killing vector for any static spherically symmetric black hole geometry and the positive frequency modes associated with this timelike Killing vector results in a vacuum state known as the Boulware vacuum state. The Boulware vacuum state is 
denoted by $|0\rangle_B$ and $|0\rangle_B\leftrightarrow|0\rangle_I$ which indicates the Boulware vacuum state is analogous to the Rindler vacuum state in region I. Another timelike Killing vector which can be defined is $-\partial_t$ and the positive frequency modes associated with this time like Killing vectors results in the $|0\rangle_{\bar{B}}$, also known as the anti-Boulware vacuum state. The anti-Boulware vacuum state is analogous to $|0\rangle_{IV}$. From the analogy among the Hartle-Hawking (Boulware, anti-Boulware) and Minkowski (Rindler I, Rindler IV) vacuum states, we can rewrite eq.(\ref{1.47}) in a static spherically black hole geometry as
\begin{equation}\label{1.66}
|0\rangle_H^{\omega_i}=\frac{1}{\cosh \sigma_{\omega_i}}\sum_n \tanh ^n\sigma_{\omega_i}|n\rangle_B^{\omega_i}|n\rangle_{\bar{B}}^{\omega_i}
\end{equation}
where $|0\rangle_H=\otimes_{j}|0\rangle_H^{\omega_j}$ and 
\begin{equation}\label{1.67}
\tanh \sigma_{\omega_i}=e^{-\frac{\pi\omega_i}{\mathcal{a}}}=\exp\left(-\frac{\pi\omega_i\sqrt{f(\mathcal{r})}}{\kappa}\right)~.
\end{equation}
The above result comes from direct analogy with the corresponding result in the Minkowski-Rindler scenario. 
%The derivation of this result is presented in an appendix.
For a quantum corrected black hole metric we can recast eq.(\ref{1.67}) as 
\begin{equation}\label{1.68}
\tanh \sigma_{\omega_i}=e^{-\frac{2\pi\omega_iGM\left(1-\frac{2GM\mathcal{r}}{\mathcal{r}^2+\tilde{\omega} G}\right)\left(GM+\sqrt{G^2M^2-\tilde{\omega}G}\right)^2}{G^2M^2+GM\sqrt{G^2M^2-\tilde{\omega}G}-\tilde{\omega}G}}~.
\end{equation}
As $\tilde{\omega}$ is a quantum gravity correction (which is very small), we can recast eq.(\ref{1.68}) in a much simpler form given by
\begin{equation}\label{1.69}
\begin{split}
\tanh\sigma_{\omega_i}\simeq e^{-4\pi\omega_iGM\sqrt{1-\frac{2GM}{\mathcal{r}}}\left(1+\frac{\tilde{\omega}}{4GM^2}+\frac{\tilde{\omega}G^2M}{\mathcal{r}^2(\mathcal{r}-2GM)}\right)}~.
\end{split}
\end{equation} 
For the quantum corrected black hole metric, we redefine the $\sigma_{\omega_i}$ term as $\mathcal{r}_{\tilde{\omega},i}$. The one particle Hartle-Hawking state takes the form given as (for a quantum corrected black hole)
\begin{equation}\label{1.70}
|1\rangle_H^{\omega_i}=\frac{1}{\cosh^2 \mathcal{r}_{\tilde{\omega},i}}\sum_{n=0}^\infty\tanh^n \mathcal{r}_{\tilde{\omega},i}\sqrt{n+1}|n+1\rangle_B^{\omega_i}|n\rangle_{\bar{B}}^{\omega_i}~.
\end{equation}
Here, we consider a maximally entangled bipartite state in the basis of an observer freely falling into the event horizon of a black hole as
\begin{equation}\label{1.71}
|\psi\rangle=\frac{1}{\sqrt{2}}\left(|0\rangle_A^{\omega_i}|0\rangle_R^{\omega_i}+|1\rangle_A^{\omega_i}|1\rangle_R^{\omega_i}\right)~.
\end{equation}
The suffix `$A$' in the first part of the system (out of the two subsystems) denotes freely falling Alice and the second subsystem is for Rob, who is at a distance
 $\mathcal{r}$ near the event horizon of the quantum corrected black hole.
\section{Logarithmic Negativity and Mutual Information}\label{IV}
\begin{center}
\begin{figure}[ht!]
\includegraphics[scale=0.35]{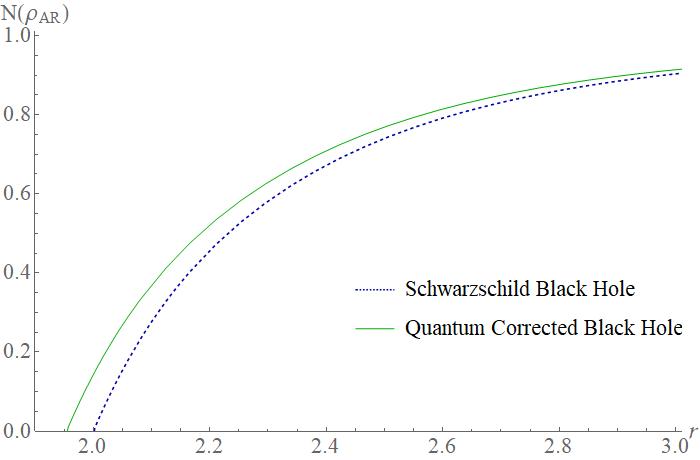}
\caption{Logarithmic negativity vs radial distance (of the observer) plot for a Schwarzschild and quantum corrected black hole.\label{Negativity_plot}}
\end{figure}
\end{center}

\begin{center}
\begin{figure}[ht!]
\includegraphics[scale=0.35]{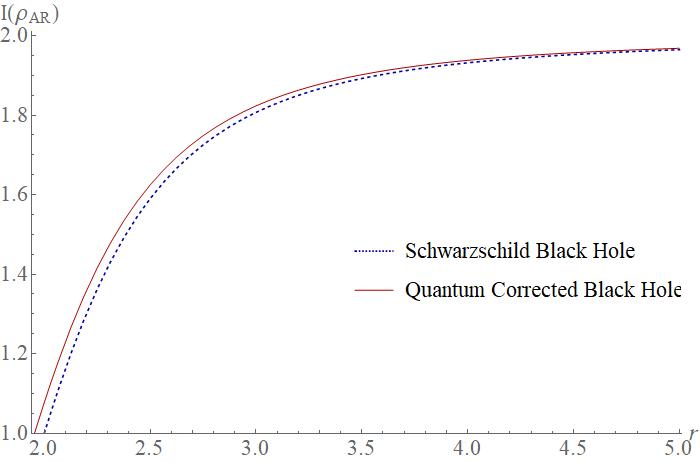}
\caption{Mutual information vs radial distance (of the observer) plot for a Schwarzschild and quantum corrected black hole.\label{Mutual_Information_plot}}
\end{figure}
\end{center}

\noindent In this section, we shall obtain the logarithmic negativity and mutual information corresponding to the maximally entangled biparite state given in eq.(\ref{1.71}). Our main aim is to do a side by side comparison for the case of a Schwarzschild and a quantum corrected black hole to truly investigate the effect of the underlying quantum nature of the black hole. Before proceeding further it is important to note that $|0\rangle_A\leftrightarrow|0\rangle_H$ and $|0\rangle_R$ for a fixed frequency value is described by eq.(\ref{1.66}). Boulware and anti-Boulware states are causally disconnected and Rob is causally disconnected from accessing the anti-Boulware states. As a result, we shall be tracing over the anti-Boulware states which shall lead to a mixed state. The reduced density matrix is given by
\begin{equation}\label{1.72}
\begin{split}
\rho_{AR}=&\sum_{m=0}^\infty~_{\bar{B}}\!\langle m|\psi\rangle\langle\psi|m\rangle_{\bar{B}}\\
=&\frac{1}{2\cosh^2\mathcal{r}_{\tilde{\omega},i}}\sum_{n=0}^\infty\tanh^{2n} \mathcal{r}_{\tilde{\omega},i}\biggr[|0~n\rangle\langle0~n|\\&+\frac{\sqrt{n+1}}{\cosh \mathcal{r}_{\tilde{\omega},i}}\left(|1~n+1\rangle\langle0~n|+|0~n\rangle\langle1~n+1|\right)\\&+\frac{n+1}{\cosh^2\mathcal{r}_{\tilde{\omega},i}}|1~n+1\rangle\langle1~n+1|\biggr]~.
\end{split}
\end{equation}
We shall now make use of the partial transpose criteria which shall provide us with sufficient criteria for entanglement. The $\{n,n+1\}$ block of the reduced density matrix in eq.(\ref{1.72}) is given by
%\begin{equation}\label{1.73}
%\left(\frac{1}{2\cosh^2\mathcal{r}_{\tilde{\omega},i}}\right)\left[\begin{tabular}{c c | c c}
%$\mathcal{B}^{0n}_{0n}|0n\rangle\langle 0n|$&$~\mathcal{B}^{0n}_{1n}|0n\rangle\langle 1n|~$&$~\mathcal{B}^{0n}_{0n+1}|0n\rangle\langle 0n+1|$&$~\mathcal{B}^{0n}_{1n+1}|0n\rangle\langle 1n+1|$\\{}&{}&{}&{}\\
%$\mathcal{B}^{1n}_{0n}|1n\rangle\langle 0n|$&$\mathcal{B}^{1n}_{1n}|1n\rangle\langle 1n|~$&$~\mathcal{B}^{1n}_{0n+1}|1n\rangle\langle 0n+1|$&$~\mathcal{B}^{1n}_{1n+1}|1n\rangle\langle 1n+1|$\\{}&{}&{}&{}\\\hline{}&{}&{}&{}\\
%$\mathcal{B}^{0n+1}_{0n}|0n+1\rangle\langle 0n|$&$~\mathcal{B}^{0n+1}_{1n}|0n+1\rangle\langle 1n|~$&$~\mathcal{B}^{0n+1}_{0n+1}|0n+1\rangle\langle 0n+1|$&$~|\mathcal{B}^{0n+1}_{1n+1}0n+1\rangle\langle 1n+1|$\\{}&{}&{}&{}\\
%$\mathcal{B}^{1n+1}_{0n}|1n+1\rangle\langle 0n|~$&$\mathcal{B}^{1n+1}_{1n}|1n+1\rangle\langle 1n|~$&$~\mathcal{B}^{1n+1}_{0n+1}|1n+1\rangle\langle 0n+1|$&$~\mathcal{B}^{1n+1}_{1n+1}|1n+1\rangle\langle 1n+1|$
%\end{tabular}\right]
%\end{equation}
\begin{equation}\label{1.73}
\left(\frac{1}{2\cosh^2\mathcal{r}_{\tilde{\omega},i}}\right)\left[\begin{tabular}{c c | c c}
$\mathcal{B}^{0n}_{0n}$&$~\mathcal{B}^{0n}_{1n}~$&$~\mathcal{B}^{0n}_{0n+1}$&$~\mathcal{B}^{0n}_{1n+1}$\\{}&{}&{}&{}\\
$\mathcal{B}^{1n}_{0n}$&$\mathcal{B}^{1n}_{1n}~$&$~\mathcal{B}^{1n}_{0n+1}$&$~\mathcal{B}^{1n}_{1n+1}$\\{}&{}&{}&{}\\\hline{}&{}&{}&{}\\
$\mathcal{B}^{0n+1}_{0n}$&$~\mathcal{B}^{0n+1}_{1n}~$&$~\mathcal{B}^{0n+1}_{0n+1}$&$~|\mathcal{B}^{0n+1}_{1n+1}$\\{}&{}&{}&{}\\
$\mathcal{B}^{1n+1}_{0n}~$&$\mathcal{B}^{1n+1}_{1n}~$&$~\mathcal{B}^{1n+1}_{0n+1}$&$~\mathcal{B}^{1n+1}_{1n+1}$
\end{tabular}\right]
\end{equation}
where $\mathcal{B}^{ab}_{cd}$ denotes the coefficient associated with the $|ab\rangle\langle cd|$ state.
After taking partial tranpose of the matrix in eq.(\ref{1.73}), we obtain the following matrix
\begin{widetext}
\begin{equation}\label{1.74}
\left(\frac{1}{2\cosh^2\mathcal{r}_{\tilde{\omega},i}}\right)\left[\begin{tabular}{c c | c c}
$\mathcal{B}^{0n}_{0n}$&$~\mathcal{B}^{0n}_{1n}~$&$~\mathcal{B}^{0n}_{0n+1}$&$~\mathcal{B}^{1n}_{0n+1}$\\{}&{}&{}&{}\\
$\mathcal{B}^{1n}_{0n}$&$\boxed{\mathcal{B}^{1n}_{1n}}~$&$~\boxed{\mathcal{B}^{0n}_{1n+1}}$&$~\mathcal{B}^{1n}_{1n+1}$\\{}&{}&{}&{}\\\hline{}&{}&{}&{}\\
$\mathcal{B}^{0n+1}_{0n}$&$~\boxed{\mathcal{B}^{1n+1}_{0n}}~$&$~\boxed{\mathcal{B}^{0n+1}_{0n+1}}$&$~|\mathcal{B}^{0n+1}_{1n+1}$\\{}&{}&{}&{}\\
$\mathcal{B}^{0n+1}_{1n}~$&$\mathcal{B}^{1n+1}_{1n}~$&$~\mathcal{B}^{1n+1}_{0n+1}$&$~\mathcal{B}^{1n+1}_{1n+1}$
\end{tabular}\right]~.
\end{equation}
The new matrix consisting of the boxed elements from eq.(\ref{1.74}) is given by
\begin{equation}\label{1.75}
\mathcal{P}_{n,n+1}=\frac{\tanh^{2n}\mathcal{r}_{\tilde{\omega},i}}{2\cosh^2\mathcal{r}_{\tilde{\omega},i}}
\begin{bmatrix}
\frac{n}{\sinh^2\mathcal{r}_{\tilde{\omega},i}}&\frac{\sqrt{n+1}}{\cosh \mathcal{r}_{\tilde{\omega},i}}\\
\frac{\sqrt{n+1}}{\cosh \mathcal{r}_{\tilde{\omega},i}}&\tanh^2\mathcal{r}_{\tilde{\omega},i}~.
\end{bmatrix}
\end{equation}
The eigenvalues of the $\mathcal{P}_{n,n+1}$ matrix are given as
\begin{equation}\label{1.76}
\begin{split}
\xi_{n,\pm}=&\frac{\tanh^{2n}\mathcal{r}_{\tilde{\omega},i}}{4\cosh^2\mathcal{r}_{\tilde{\omega},i}}\Biggr[\left(\tanh^2\mathcal{r}_{\tilde{\omega},i}+\frac{n}{\sinh^2\mathcal{r}_{\tilde{\omega},i}}\right)\pm\sqrt{\left(\tanh^2\mathcal{r}_{\tilde{\omega},i}+\frac{n}{\sinh^2\mathcal{r}_{\tilde{\omega},i}}\right)^2+\frac{4}{\cosh^2\mathcal{r}_{\tilde{\omega},i}}}\Biggr]~.
\end{split}
\end{equation}
From eq.(\ref{1.76}), it is straightforward to infer that $\xi_{n,-}<0$. The logarithmic negativity is obtained as
\begin{equation}\label{1.77}
\begin{split}
&N(\rho_{AR})=\log_2||\rho_{AR}^T||\\
=&\log_2\left[1+\sum_{n=0}^\infty\left(|\xi_{n,-}| -\xi_{n,-}\right)\right]\\
=&\log_2\left[1-2\sum_{n=0}^\infty\xi_{n,-}\right]\\
=&\log_2\biggr[1+\sum_{n=0}^\infty\frac{\tanh^{2n} \mathcal{r}_{\tilde{\omega},i}}{2\cosh^2\mathcal{r}_{\tilde{\omega},i}}\sqrt{\left(\tanh^2\mathcal{r}_{\tilde{\omega},i}+\frac{n}{\sinh^2\mathcal{r}_{\tilde{\omega},i}
}\right)^2+\frac{4}{\cosh^2\mathcal{r}_{\tilde{\omega},i}}}-\sum_{n=0}^\infty\frac{\tanh^{2n}\mathcal{r}_{\tilde{\omega},i}}{2\cosh^2\mathcal{r}_{\tilde{\omega},i}}\left(\frac{n}{\sinh^2\mathcal{r}_{\tilde{\omega},i}}+\tanh^2\mathcal{r}_{\tilde{\omega},i}\right)\biggr]\\
=&\log_2\biggr[\frac{1}{2\cosh^2\mathcal{r}_{\tilde{\omega},i}}+\sum_{n=0}^\infty\frac{\tanh^{2n} \mathcal{r}_{\tilde{\omega},i}}{2\cosh^2\mathcal{r}_{\tilde{\omega},i}}\sqrt{\left(\tanh^2\mathcal{r}_{\tilde{\omega},i}+\frac{n}{\sinh^2\mathcal{r}_{\tilde{\omega},i}
}\right)^2+\frac{4}{\cosh^2\mathcal{r}_{\tilde{\omega},i}}}\biggr]\\
=&\log_2\left[\frac{1}{2\cosh^2\mathcal{r}_{\tilde{\omega},i}}+\Lambda(\mathcal{r}_{\tilde{\omega},i})\right]
\end{split}
\end{equation}
where
\begin{equation}\label{1.78}
\Lambda(\mathcal{r}_{\tilde{\omega},i})=\sum_{n=0}^\infty\frac{\tanh^{2n} \mathcal{r}_{\tilde{\omega},i}}{2\cosh^2\mathcal{r}_{\tilde{\omega},i}}\sqrt{\left(\tanh^2\mathcal{r}_{\tilde{\omega},i}+\frac{n}{\sinh^2\mathcal{r}_{\tilde{\omega},i}
}\right)^2+\frac{4}{\cosh^2\mathcal{r}_{\tilde{\omega},i}}}~.
\end{equation}
\end{widetext}
For an observer at infinite distance, $a(\mathcal{r}\rightarrow\infty)=0$ leading to $N(\rho_{AR})=1$. When the observer is on the event horizon of the black hole then $a(r_+)\rightarrow\infty$ which is identical to the condition $\mathcal{r}_{\tilde{\omega},i}\rightarrow\infty$. To obtain the value of the entanglement negativity at this point, we need to truly investigate the bound on the negativity value in this limit. 
It is straightforward to obtain a bound on the summation term in the above equation. We know that $a^2+b^2<(a+b)^2$ and we can write down the following inequality
\begin{equation}\label{1.79}
\begin{split}
\Lambda(\mathcal{r}_{\tilde{\omega},i})&<\sum_{n=0}^\infty\frac{\tanh^{2n} \mathcal{r}_{\tilde{\omega},i}}{2\cosh^2\mathcal{r}_{\tilde{\omega},i}}\biggr[\tanh^2\mathcal{r}_{\tilde{\omega},i}+\frac{n}{\sinh^2\mathcal{r}_{\tilde{\omega},i}}\\&+\frac{2}{\cosh\mathcal{r}_{\tilde{\omega},i}}\biggr]\\
&=\frac{1}{2}\left(1+\frac{2}{\cosh\mathcal{r}_{\tilde{\omega},i}}+\tanh^2\mathcal{r}_{\tilde{\omega},i}\right)\\&<1+\frac{1}{\cosh\mathcal{r}_{\tilde{\omega},i}}~.
\end{split}
\end{equation}
Now in the $\mathcal{r}_{\tilde{\omega},i}\rightarrow\infty$ limit $\Lambda$ goes to 1. Hence, in the infinite acceleration case or when the observer is on the event horizon of the black hole, the logarithmic negativity becomes 0. In the $\tilde{\omega}\rightarrow0$ limit $\mathcal{r}_{\tilde{\omega},i}\rightarrow \mathcal{r}_{Sch.,i}$ with ``$Sch.$" denotes the case for a Schwarzschild black hole. For the next part of our analysis, we shall be denoting $\mathcal{r}_{\tilde{\omega},i}$ as $\mathcal{R}_i$. The logarithmic negativity from eq.(\ref{1.77}) can be expressed in terms of $\mathcal{R}_i$ as 
\begin{equation}\label{1.80}
\begin{split}
&N(\rho_{AR})\simeq\log_2\biggr[\frac{1}{2\cosh^2\mathcal{R}_i}\left(1+\tilde{\omega}\mathcal{K}_i\sinh^2\mathcal{R}_i\right)+\\&\sum_{n=0}^\infty\frac{\tanh^{2n}\mathcal{R}_i}{2\cosh^2\mathcal{R}_i}\sqrt{\left(\frac{n}{\sinh^2\mathcal{R}_i}+\tanh^2\mathcal{R}_i\right)^2+\frac{4}{\cosh^2\mathcal{R}_i}}\biggr]\\&\times\Biggr(1+\tilde{\omega}\mathcal{K}_i\biggr(\sinh^2\mathcal{R}_i-n+\biggr[2\tanh^2\mathcal{R}_i+\biggr(\frac{n}{\sinh^2\mathcal{R}_i}\\&+\tanh^2\mathcal{R}_i\biggr)\left[\frac{n}{\sinh^2\mathcal{R}_i}+n-\tanh^2\mathcal{R}_i\right]\biggr]\biggr/\biggr[\biggr(\frac{n}{\sinh^2\mathcal{R}_i}\\&+\tanh^2\mathcal{R}_i\biggr)^2+\frac{4}{\cosh^2\mathcal{R}_i}\biggr]\biggr)\Biggr)
\end{split}
\end{equation}
where 
\begin{equation}\label{1.81}
\mathcal{K}_i\equiv 8\pi\omega_iGM\sqrt{1-\frac{2GM}{\mathcal{r}}}\left[\frac{1}{4GM^2}+\frac{G^2M}{\mathcal{r}^2(\mathcal{r}-2GM)}\right]~.
\end{equation}
Eq.(\ref{1.80}) is one of the main results in our paper. In order to plot Fig.(\ref{Negativity_plot}), we set $G=0.1\mathcal{l}_0^2$, $M=1.0 \mathcal{l}_0^{-1}$, $\omega_i=\frac{1}{\pi}\mathcal{l}_0^{-1}$, and $\tilde{\omega}=0.9$ with respect to some arbitrary length scale $\mathcal{l}_0$. Here, the values are chosen in a manner such that the quantum effects get amplified. From Fig.(\ref{Negativity_plot}), we observe that the negativity has a slower rate of decreasing than the Schwarzschild black hole. Hence, if an observer finds out that at the Schwarzschild radius, the logarithmic negativity does not drop to zero, then it will be a direct detection of the quantum nature of the black hole. It is also important to note that the negativity goes to zero on the event horizon radius of the black hole (the points for each curve where they meet on the $N(\rho_{AR})=0$ axis) which signifies that the states does nit possess any distillable entanglement anymore.
It is important to note that we have made use of eq.(\ref{1.68}) instead of eq.(\ref{1.69}) to obtain Fig.(\ref{Negativity_plot}) (and later Fig.(s)(\ref{Mutual_Information_plot},\ref{Fidelity_plot})).

\noindent Our next aim is to calculate the mutual information and compare between the Shwarzschild and quantum corrected case. The mutual information gives one the idea of the total amount of correlation. The mutual information is given by
\begin{equation}\label{1.82}
I(\rho_{AR})=S(\rho_A)+S(\rho_{R})-S(\rho_{AR})
\end{equation}
where $S(\rho)=-\text{tr}\left(\rho\log_2\rho\right)=-\sum_{n}\rho_{n,n}\log_2\rho_{n,n}$. In eq.(\ref{1.82}), $\rho_A$ denotes Alice's density matrix while Rob's states are traced out. The values of the individual entropies can be obtained as follows
\begin{align}
S(\rho_A)=&1~,\label{1.83}\\
S(\rho_R)=&-\sum_{n=0}^\infty\frac{\tanh^{2n}\mathcal{r}_{\tilde{\omega},i}}{2\cosh^2\mathcal{r}_{\tilde{\omega,i}}}\left(1+\frac{n}{\sinh^2\mathcal{r}_{\tilde{\omega,i}}}\right)\label{1.84}\\&\times\log_2\biggr[1\nonumber+\frac{n}{\sinh^2\mathcal{r}_{\tilde{\omega,i}}}\biggr]~,\\
S(\rho_{AR})=&-\sum_{n=0}^\infty\frac{\tanh^{2n}\mathcal{r}_{\tilde{\omega},i}}{2\cosh^2\mathcal{r}_{\tilde{\omega,i}}}\left(1+\frac{n+1}{\cosh^2\mathcal{r}_{\tilde{\omega,i}}}\right)\label{1.85}\\&\times\log_2\biggr[1+\frac{n+1}{\cosh^2\mathcal{r}_{\tilde{\omega,i}}}\biggr]~.\nonumber
\end{align}
\begin{widetext}
Substituting eq.(s)(\ref{1.83}-\ref{1.85}) in eq.(\ref{1.82}), we obtain the following relation
\begin{equation}\label{1.86}
\begin{split}
I(\rho_{AR})=&1-\sum_{n=0}^\infty\frac{\tanh^{2n}\mathcal{r}_{\tilde{\omega},i}}{2\cosh^2\mathcal{r}_{\tilde{\omega,i}}}\left[\left(1+\frac{n}{\sinh^2\mathcal{r}_{\tilde{\omega,i}}}\right)\log_2\left[1+\frac{n}{\sinh^2\mathcal{r}_{\tilde{\omega,i}}}\right]-\left(1+\frac{n+1}{\cosh^2\mathcal{r}_{\tilde{\omega,i}}}\right)\log_2\left[1+\frac{n+1}{\cosh^2\mathcal{r}_{\tilde{\omega,i}}}\right]\right]\\
\simeq& 1-\sum_{n=0}^\infty\frac{\tanh^{2n}\mathcal{R}_i}{2\cosh^2\mathcal{R}_i}\left(1+\tilde{\omega}\mathcal{K}_i(\sinh^2\mathcal{R}_i-n)\right)\Biggr[\left(1+\frac{n}{\sinh^2\mathcal{R}_i}\left(1+\tilde{\omega}\mathcal{K}_i(1+\sinh^2\mathcal{R}_i)\right)\right)\log_2\biggr[1+\frac{n}{\sinh^2\mathcal{R}_i}\\\times&\left(1+\tilde{\omega}\mathcal{K}_i(1+\sinh^2\mathcal{R}_i)\right)\biggr]-\left(1+\frac{n+1}{\cosh^2\mathcal{R}_i}\left(1+\tilde{\omega}\mathcal{K}_i\sinh^2\mathcal{R}_i\right)\right)\log_2\left[1+\frac{n+1}{\cosh^2\mathcal{R}_i}\left(1+\tilde{\omega}\mathcal{K}_i\sinh^2\mathcal{R}_i\right)\right]\Biggr]~.
\end{split}
\end{equation}
Eq.(\ref{1.86}) is also one of the main results in our paper. 
\end{widetext}
We shall now investigate the entanglement-degradation for a quantum corrected black hole and compare it with that of the Schwarzschild black hole.
Using the same parameters as before, we plot mutual information vs $\mathcal{r}$ in Fig.(\ref{Mutual_Information_plot}). In order to obtain Fig.(\ref{Mutual_Information_plot}), we have used the value of $\mathcal{r}_{\tilde{\omega}_i}$ from eq.(\ref{1.68}) ($\sigma_{\omega_i}=\mathcal{r}_{\tilde{\omega}_i}$ in this equation) instead of eq.(\ref{1.69}). It is straightforward to observe that the entangement degradation gets significant as the observer approaches the event horizons of the respective black holes. It is again important to notice that for the quantum corrected black the mutual information degrades at a slower rate. When the mutual information becomes 1, there are no distillable entanglement left between the two states. It can be a direct detection of quantum gravity signatures if for the observer being at the Schwarzschild radius, the mutual information doesn't go to unity which implies that the entanglement does not degrade completely. It is very difficult to costruct such experimental scenarios where the degradation in mutual information is directly observed but it may be possible to do the same in future with advanced experimental set ups. In the next section, we shall demonstrate the entire set up as a quantum channel with a completely positive and trace preserving map and try to obtain the entanglement fidelity for the same channel.
\section{Noisy quantum channel and entanglement fidelity}\label{V}
\noindent We start with the initial density matrix (anti-Boulware states traced out) 
\begin{equation}\label{1.87}
\begin{split}
\rho_{AR}^\mathcal{I}&=|\phi\rangle\langle\phi|\\
&=\frac{1}{2}\left(|00\rangle\langle 00|+|00\rangle\langle11|+|11\rangle\langle00|+|11\rangle\langle11|\right)
\end{split}
\end{equation}
where $|\phi\rangle=\frac{1}{\sqrt{2}}(|00\rangle+|11\rangle)$. We need to construct a map such that we obtain $\rho_{AR}$ in eq.(\ref{1.72}) from the above equation. We consider a map of the following form \cite{Schumacher}
\begin{equation}\label{1.88}
\rho_{AR}=\mathcal{E}\left(\rho_{AR}^\mathcal{I}\right)=\sum_n\mathcal{S}_n\rho_{AR}^{\mathcal{I}}\mathcal{S}_n^\dagger=\sum_n\mathcal{S}_n|\phi\rangle\langle\phi|\mathcal{S}_n^\dagger~.
\end{equation}
One can obtain the analytical form of $\mathcal{S}_n$ as
\begin{equation}\label{1.89}
\mathcal{S}_n=\frac{1}{\sqrt{n!}}\frac{\tanh^n\mathcal{r}_{\tilde{\omega},i}}{\cosh\mathcal{r}_{\tilde{\omega},i}}(\text{sech}\mathcal{r}_{\tilde{\omega},i})^{\hat{N}_A}\otimes\left(\hat{a}_B^\dagger\right)^n
\end{equation}
with $\hat{N}_A$ being the number opertor whose action is defined on Alice's Hilbert space (Hawking-Hartle states) and $\hat{a}_B^\dagger$ being the raising operator for the states measured by Rob (Boulware states). Now the operator $\mathcal{S}_n$  is an operator of the Hilbert space where the density matrix $\rho^\mathcal{I}_{AR}$ is prepared. Hence the map $\mathcal{E}$ is a positive map \cite{Schumacher, Krauss}. It is also straightforward to check that $\text{tr}(\rho^{\mathcal{I}}_{AR})=\text{tr}(\rho_{AR})$. Hence, the map $\mathcal{E}$ is a CPTP map. Our final aim is to investigate how this quantum channel preserves the initial entanglement. For this we need to calculate entanglement fidelity given by \cite{Schumacher}
\begin{equation}\label{1.90}
\begin{split}
\mathcal{F}_\mathcal{E}=\sum_{n=0}^{\infty}\text{tr}\left[\rho_{AR}^\mathcal{I}\mathcal{S}_n\right]\text{tr}\left[\rho_{AR}^\mathcal{I}\mathcal{S}_n^\dagger\right]~.
\end{split}
\end{equation}
The analytical forms of the two traces are given by
\begin{equation}\label{1.91}
\begin{split}
\text{tr}\left[\rho_{AR}^\mathcal{I}\mathcal{S}_n\right]&=\frac{\tanh^n\mathcal{r}_{\tilde{\omega},i}}{2\cosh\mathcal{r}_{\tilde{\omega},i}}\left(1+\frac{\sqrt{n+1}}{\cosh\mathcal{r}_{\tilde{\omega},i}}\right)\delta_{n,0}\\
&=\frac{1}{2\cosh\mathcal{r}_{\tilde{\omega},i}}\left(1+\frac{1}{\cosh\mathcal{r}_{\tilde{\omega},i}}\right)\delta_{n,0}\\
&=\text{tr}\left[\rho_{AR}^\mathcal{I}\mathcal{S}_n^\dagger\right]~.
\end{split}
\end{equation} 
Using eq.(\ref{1.91}) in eq.(\ref{1.90}), we obtain the entanglement fidelity given by
\begin{equation}\label{1.92}
\begin{split}
\mathcal{F}_\mathcal{E}&=\sum_{n=0}^\infty\text{tr}\left[\rho_{AR}^\mathcal{I}\mathcal{S}_n\right]\text{tr}\left[\rho_{AR}^\mathcal{I}\mathcal{S}_n^\dagger\right]\\&=\sum_{n=0}^\infty\frac{\tanh^{2n}\mathcal{r}_{\tilde{\omega},i}}{4\cosh^2\mathcal{r}_{\tilde{\omega},i}}\left(1+\frac{\sqrt{n+1}}{\cosh\mathcal{r}_{\tilde{\omega},i}}\right)^2(\delta_{n,0})^2\\&=\frac{1}{4\cosh^2\mathcal{r}_{\tilde{\omega},i}}\left(1+\frac{1}{\cosh\mathcal{r}_{\tilde{\omega},i}}\right)^2\\
&\simeq\mathcal{F}_\mathcal{E}^{Sch.}\left(1+\tilde{\omega}\mathcal{K}_i\sinh^2\mathcal{R}_i\left(1+\frac{1}{1+\cosh\mathcal{R}_i}\right)\right)~.
\end{split}
\end{equation}
where
\begin{equation}\label{1.93}
\mathcal{F}_{\mathcal{E}}^{Sch.}=\frac{1}{4\cosh^2\mathcal{R}_i}\left(1+\frac{1}{\cosh\mathcal{R}_i}\right)^2
\end{equation}
denotes the entanglement fidelity for a Schwarzschild black hole.
\begin{center}
\begin{figure}[ht!]
\includegraphics[scale=0.35]{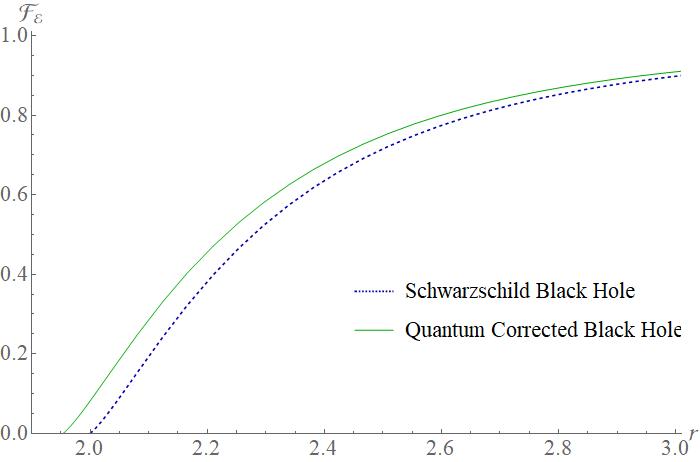}
\caption{Entanglement fidelity vs $\mathcal{r}$ plot for a Schwarzschild and a quantum corrected black hole.\label{Fidelity_plot}}
\end{figure}
\end{center}
In Fig.(\ref{Fidelity_plot}), we plot the entanglement fidelity vs the distance of the observer from $\mathcal{r}=0$. It is important to observe from Fig.(\ref{Fidelity_plot}) that near the event horizon of the black hole (which depicts the infinite acceleration limit in the flat spacetime case), the entanglement fidelity approaches zero and the rate of degradation is slower for the quantum corrected black hole (compared to the Schwarzschild black hole) which shows a similar behaviour as shown by logarithmic negativity and mutual information.
\section{Conclusion}\label{VI}
\noindent We investigate the phenomenon of entanglement degradation for a quantum corrected black hole, in the vicinity of the event horizon of the same. We observe that in the near horizon approximation, it is possible to write down any static and spherically symmetric metric in a Rindler form which helps later in identifying three time-like Killing vectors and ultimately in identifying the vacuum modes and their analogy with the flat spacetime case. For the next part of our analysis, we obtain the logarithmic negativity for the quantum corrected black hole using the partial transpose criterion of the reduced density matrix and expressed it in terms of the Schwarzschild parameters. Then we have plotted logarithmic negativity with respect to the change in the position of the observer $\mathcal{r}$ for a quantum corrected black hole and compared it with that of the Schwarzshild black hole. We observe that the logarithmic negativity asymptotically reaches unity when the observer is sitting very far away from each of the black holes and attains a zero value for an observer sitting on the event horizon of the black hole. It is although important to note that when the logarithmic negativity reaches zero value for observer sitting at the event horizon radius of the Schwarzschild black hole, it still would have been non zero if there are underlying quantum gravity corrections in the black hole. Next we have calculated the mutual information for the Alice-Rob bipartite state for the quantum corrected black hole. We have then plotted the mutual information with respect to $\mathcal{r}$ for both the black holes and observe that very near the event horizon radius, mutual information drops from 2 to very close to unity and reaches unity while the observer is sitting on the event horizon of the black hole. Similar to previous case the mutual information has a slower rate of fall for the quantum corrected black hole. It affirms that if a black hole has underlying quantum gravity corrections (which is almost impossible to notice for any observer) then even if the observer is at the Schwarzschild radius there will still be some distillable entanglement left. This observation may be considered as an important quantum gravity signature which can be looked for. Finally, we consider the entire procedure as a quantum channel and obtained a completely positive trace preserving map which translates the initial stationary entangled state to a mixed state in the black hole spacetime. We finally calculate the entanglement fidelity to investigate how the quantum channel preserves entanglement. We find out that the entanglement fidelity degrades near the vicinity of the event horizon of the black hole and as per the earlier cases the rate of fall is slower in case of the quantum corrected black hole. It is then important to conclude that quantum gravity corrections delay entanglement degradation and the interesting physics occurs in the vicinity of the event horizon of the black hole. Our future plan involves doing a rigorous calculation of the Bogoliubov coefficients and obtain the Hawking-Hartle and Boulware vacuum connection considering the effects of the curved background rather by using an analogy. 

\end{document}